\documentclass[12pt]{article}

\usepackage{amssymb}
\usepackage[dvips]{graphicx}
\input epsf

\def\appendix#1{
  \addtocounter{section}{1}
  \setcounter{equation}{0}
  \renewcommand{\thesection}{\Alph{section}}
  \section*{Appendix \thesection\protect\indent \parbox[t]{11.715cm} {#1}}
  \addcontentsline{toc}{section}{Appendix \thesection\ \ \ #1}
  }

\newcommand {\bd}{\begin{displaymath}}
\newcommand {\ed}{\end{displaymath}}
\newcommand {\eq}{\begin{equation}}
\newcommand {\beq}{\begin{equation}}
\newcommand {\eeq}{\end{equation}}
\newcommand {\beqa}{\begin{eqnarray}}
\newcommand {\eeqa}{\end{eqnarray}}
\newcommand {\n}{\nonumber \\}
\newcommand {\tr}{{\rm tr\,}}
\newcommand {\Tr}{\mbox{Tr\,}}

\newcommand {\ee}{\mbox{e}}
\newcommand {\latsp}{\epsilon}

\newcommand {\dd}{\mbox{d}}

\newcommand {\del}{\partial}

\newcommand {\defeq}{\stackrel{\rm def}{=}}

\font\mybb=msbm10 at 12pt
\def\bb#1{\hbox{\mybb#1}}

\def\IR{{\bb R}}
\def\IZ{{\bb Z}}

\newcommand{\id}{{1\!\!1}} 

\begin{document}

\setlength{\oddsidemargin}{0cm}
\setlength{\baselineskip}{7mm}

\begin{titlepage}

\baselineskip=14pt

\renewcommand{\thefootnote}{\fnsymbol{footnote}}
\begin{normalsize}
\begin{flushright}
\begin{tabular}{l}
hep-th/0107110\\
\hfill{ }\\
July 2001
\end{tabular}
\end{flushright}
  \end{normalsize}



\vspace{1cm}

\vspace*{0cm}
    \begin{Large}
       \begin{center}
{Noncommutative Chiral Gauge Theories on the Lattice}\\
{with Manifest Star-Gauge Invariance}\\

       \end{center}
    \end{Large}
\vspace{1cm}

\begin{center}
J. N{\sc ishimura}\footnote{
On leave from Department
of Physics, Nagoya University, Nagoya 464-8602, Japan,\\
e-mail address : nisimura@nbi.dk}
\setcounter{footnote}{2}
           {\sc and}
M.A. V{\sc \'azquez-Mozo}\footnote{e-mail address : vazquez@nbi.dk} \\

      \vspace{1cm}
         {\it The Niels Bohr Institute\\ Blegdamsvej 17 \\ 
          DK-2100 Copenhagen \O, Denmark}\\[4mm]

\end{center}

\vskip 1 cm

\hspace{5cm}

\begin{abstract}
\noindent
We show that 
noncommutative U($r$) gauge theories with a chiral
fermion in the adjoint representation
can be constructed on the lattice with manifest star-gauge invariance
in arbitrary even dimensions.
Chiral fermions are implemented
using a Dirac operator which satisfies the Ginsparg-Wilson relation.
A gauge-invariant integration measure for the fermion fields
can be given explicitly, which simplifies the construction as compared
with
lattice chiral gauge theories in ordinary (commutative) space-time.
Our construction includes the cases where continuum calculations
yield a gauge anomaly.
This reveals a certain regularization dependence, which is
reminiscent of  parity anomaly 
in commutative space-time with odd dimensions.
We speculate that the gauge anomaly obtained 
in the continuum calculations in the present cases
can be cancelled by an appropriate counterterm.

\end{abstract}
\vfill
\end{titlepage}
\vfil\eject
\setcounter{footnote}{0}

\renewcommand{\thefootnote}{\arabic{footnote}}

\baselineskip=18pt


\section{Introduction}

\setcounter{equation}{0}
\renewcommand{\thefootnote}{\arabic{footnote}} 

Recently gauge theories on noncommutative space-time have attracted
much attention because of their intimate relationships to 
string theories (for a review see \cite{dougnek}).
In particular, 
Matrix Theory \cite{BFSS} and the IIB matrix model \cite{IKKT},
which are conjectured to provide nonperturbative definitions of string/M
theories, give rise to noncommutative gauge theory on toroidal
compactifications \cite{CDS}. The particular noncommutative toroidal
compactification is interpreted as being the result of the presence of a
background Neveu-Schwarz two-form field, and it can also be understood in the
context of open string quantization in D-brane backgrounds \cite{SW}.
Furthermore, in Ref.~\cite{AIIKKT} it has been shown that the IIB matrix model
with D-brane backgrounds is described by noncommutative gauge theory.
Based on these developments, a lattice formulation of 
noncommutative gauge theory has been 
established \cite{Bars:2000av,AMNS1,AMNS2,AMNS3}.
(see also \cite{Makeenko:2000tc,Szabo:2001qd} for reviews.)
In this paper, we apply this lattice formulation to
studying chiral fermions in noncommutative space-time.

Construction of chiral gauge theories on the lattice was considered
to be difficult for a long time 
due to the well-known no-go theorem \cite{Nielsen:1981rz}.
The situation has changed drastically since
a manifestly gauge-invariant construction of 
Abelian lattice chiral gauge theories 
has been established \cite{Luescher_abelian}.
The no-go theorem is circumvented by requiring the Dirac action
on the lattice (without species doublers) to be manifestly invariant
under a {\em modified} chiral transformation \cite{ML},
which reduces to the usual chiral 
transformation only in the continuum limit.
Such a Dirac action can be constructed
by the use of a Dirac operator \cite{overlapDirac}
which satisfies the Ginsparg-Wilson relation \cite{GW}.
The Dirac operator breaks ultra-locality, but it still preserves 
locality in the sense that the corresponding kernel decays
exponentially at long distances \cite{Hernandez:1999et}.
In this formalism 
the construction of chiral gauge theories boils down to the question
of choosing an integration measure for the chiral fermion fields
in a gauge invariant way
\footnote{While this paper was being
prepared, we received a preprint \cite{Kikukawa:2001mw}
where a different type of manifestly gauge invariant construction of 
lattice chiral gauge theories has been proposed.}.

We take the same approach in the noncommutative space-time
with arbitrary even dimensions.
When the chiral fermion transforms as the adjoint representation
under star-gauge transformations,
we can construct a star-gauge invariant 
fermion measure {\em explicitly}
by exploiting a peculiar property of noncommutative geometry.
This is a considerable simplification compared with
the construction of Abelian lattice chiral gauge theories 
in commutative space-time \cite{Luescher_abelian},
where only the {\em existence} of a gauge-invariant fermion measure 
is known nonperturbatively.
Moreover, our construction can be extended to the
case with higher-rank gauge groups in a straightforward manner.

Noncommutative chiral gauge theories have been also studied
in the continuum by various regularizations. 
In Ref.~\cite{Gracia-Bondia:2000pz} 
the gauge anomaly has been calculated
using a variant of Fujikawa's method \cite{Fujikawa,Alvarez-Gaume:1984cs}.
It was found there that four-dimensional noncommutative chiral gauge theories 
with fermions
in the fundamental representation of U($r$) are anomalous. 
The result was confirmed from the viewpoint
of descent equations \cite{Bonora:2000he},
and also reproduced diagrammatically using dimensional 
regularization \cite{Martin:2000qf}.
On the other hand, if the chiral fermions transform 
in the adjoint representation
of the gauge group, the total anomaly cancels in four dimensions 
and gauge invariance is preserved. 
This is not so suprising since four-dimensional chiral fermions in the adjoint representation 
can be formulated alternatively in terms of Majorana fermions. However this is not the
case in two dimensions, where we find that a similar calculation
yields a nonvanishing gauge anomaly.
Comparing these results with our lattice construction,
we observe a certain regularization dependence,
which is reminiscent of the parity anomaly in commutative space-time
with an odd number of dimensions  
(See Refs.\ \cite{oddD} and references therein).
We speculate that the gauge anomaly obtained in the continuum 
calculations might be cancelled by an appropriate counterterm 
in these cases.

The rest of the paper is organized as follows.
In Section \ref{matrixmodel}, we review the lattice formulation
of noncommutative gauge theories.
In Section \ref{chiral}, we introduce chiral fermion in noncommutative
gauge theories.
In Section \ref{continuum}, we present some calculations of the gauge anomaly 
in noncommutative chiral gauge theories in the continuum. Finally,
Section \ref{summary} is devoted to summary and discussions.

\setcounter{equation}{0}
\section{Lattice formulation of noncommutative gauge theories}
\label{matrixmodel}

In this section, we briefly review
the lattice formulation of
noncommutative gauge theory \cite{AMNS1,AMNS2,AMNS3}.
For concreteness, we shall work with a specific example
based on twisted Eguchi-Kawai model \cite{EK,GO}.

\subsection{discrete noncommutative geometry}
\label{discreteNC}

The starting point of noncommutative geometry (on $\IR ^{D}$)
is to replace the space-time coordinates
by hermitian operators obeying the commutation
relation
\beq
[\hat{x}_\mu,\hat{x}_\nu] = i\,\theta_{\mu\nu} \ ,
\label{Xcommutation}
\eeq
where $\theta_{\mu\nu}=-\theta_{\nu\mu}$ are real-valued
c-numbers with dimensions of length squared.
One may also introduce an anti-hermitian derivation $\hat{\del}_\mu$ 
through the commutation relation
\beq
\left[\hat{\del}_\mu\,,\, \hat{x} _\nu \right]
= \delta _ {\mu\nu} 
~~~~~~,~~~~~~\left[\hat\del_\mu\,,\,\hat\del_\nu
\right]=i\,c_{\mu\nu} \ ,
\label{deldelCR}
\eeq
where $c_{\mu\nu}=-c_{\nu\mu}$ are real-valued c-numbers.
Let us introduce a $D$-dimensional discretized torus
defined by\footnote{Since we will have to restrict $L$ to be odd later,
we already assume it at this point.}
\beq
\Lambda _\ell  = \Bigl\{  (x_1 , x_2 , \cdots , x_D)  \, | \,
x_\mu = \latsp n_\mu \,, \, n_\mu \in \IZ \, , \, 
-\frac{L-1}{2} \le n_\mu \le \frac{L-1}{2} \Bigr\} \ ,
\eeq
where $\latsp$ is the lattice spacing.
We denote the dimensionful extent of the torus by 
$\ell = \latsp L$.
One can define a discretized noncommutative torus
by constructing operators $Z_\mu$ and $\Gamma_\mu$, which
corresponds to $\ee ^{2 \pi i \hat{x}_\mu/\ell }$ and
$\ee ^{\latsp \hat{\del}_\mu }$.
These operators satisfy the commutation relations
\beqa
\label{Zprimelatticecomm}
Z_ \mu Z_ \nu
&=&  \ee ^{- 2 \pi i \Theta  _{\mu\nu}}\,Z_ \nu Z_ \mu  \\
\Gamma_\mu\, Z_ \nu \, \Gamma_\mu ^\dag 
&=&
\ee^{ 2\pi i \latsp \delta_{\mu\nu} / \ell} \, 
Z_ \nu  
\label{Zprimelatticecomm2} 
\\
\Gamma _\mu \Gamma _\nu 
&=& {\cal Z}_{\mu\nu} \Gamma _\nu \Gamma _\mu \ ,
\label{GamCR}
\eeqa
where $\Theta _{\mu\nu} = 2 \pi \theta _{\mu\nu}/\ell^2$ is the
dimensionless noncommutativity parameter
and the phase factors ${\cal Z}_{\mu\nu} = {\cal Z}_{\nu\mu}^*$
are related to $c_{\mu\nu}$ as 
${\cal Z}_{\mu\nu} = \ee ^{i \latsp ^2 c_{\mu\nu}}$.

In what follows we construct a representation of operators
$Z_\mu$ and $\Gamma_\mu$ in terms of $N \times N$ unitary matrices.
Let us start with the $Z_\mu$'s.
The commutation (\ref{Zprimelatticecomm}) is invariant under
\beq
Z_\mu \mapsto \hat{g} \, Z_\mu \, \hat{g}^\dag \ ,
\label{SUNsym}
\eeq
where $\hat{g}$ is an SU($N$) matrix and the U(1)$^D$ symmetry
\beq
Z_\mu \mapsto \ee ^{i \alpha_\mu} Z_\mu  \ .
\label{U1sym}
\eeq
We assume that $D$ is even, since we are going to introduce chiral fermions.
The form of the noncommutativity matrix is taken to be
$\Theta _{\mu\nu} = \frac{b}{L}\varepsilon _{\mu\nu}$,
where $b$ is an integer,
and $\varepsilon _{\mu\nu}$ 
is a skew-diagonal antisymmetric matrix defined by
\beq
\varepsilon
=\pmatrix{0&-1 & & & \cr 1&0& & & \cr & &\ddots& & 
\cr & & &0&-1\cr & &
& 1 &0\cr} \ .
\label{Qdiag}
\eeq
We also assume that 
$b$ and $L$ are co-prime,
and $L$ divides the dimension $N$ of the representation.
Then the necessary and sufficient condition for the existence of a
unique solution to (\ref{GamCR}) 
---up to the symmetries (\ref{SUNsym}) and (\ref{U1sym})---
is \cite{twisteater}
\beq
N=  L^{D/2} \ ,
\label{tildep0def}
\eeq
which we assume in what follows.
An explicit solution to (\ref{Zprimelatticecomm}) is given 
by \cite{twisteater}
\beqa
Z_{2j-1}&=&\id_{L}\otimes\cdots\otimes 
V_{L}
\otimes\cdots\otimes\id_{L}\n
Z_{2j} 
&=&\id_{L}\otimes\cdots\otimes 
\left(W_{L} \right)^{b}
\otimes\cdots\otimes\id_{L} 
\label{Zdef} 
\eeqa
for $j=1,\dots,D/2$.
The SU$(L)$ matrices $V_L$ and $W_L$, which appear in the $j$-th
entry of the tensor products in (\ref{Zdef}), 
are the shift and clock matrices
\beq
V_L=\pmatrix{0&1& & &0\cr &0&1& & \cr& 
&\ddots&\ddots& \cr& & &\ddots&1\cr 1& &
& &0\cr}~~~~~~,~~~~~~
W_L=\pmatrix{1& & & & \cr &\ee^{2\pi i/L}& & & \cr&
&\ee^{4\pi i/L}& & \cr& & &\ddots& \cr & & & &\ee^{2\pi i(L-1)/L}\cr}
\label{clockshift}
\eeq
obeying $V_L W_L=\ee^{2\pi i/L}\,W_L V_L$.
Since $b$ and $L$ are co-prime,
there exists a set of integers $q$ and $s$ such that
\beq
b q + L s  =1 \ .
\label{apbqpi}
\eeq
We may construct a representation of $\Gamma_{\mu}$'s as
\beqa
\Gamma_{2j-1}
&=&\id_{L}\otimes\cdots\otimes \left( W_{L} \right)^{b q}
 \otimes\cdots\otimes\id_{L} \n
\Gamma_{2j} 
&=&\id_{L}\otimes\cdots\otimes\left(V_{L} ^\dag \right)^{ q}
\otimes\cdots\otimes\id_{L} \ ,
\label{twisteat}
\eeqa
which obeys the commutation relation (\ref{Zprimelatticecomm2}) 
and (\ref{GamCR}) due to (\ref{apbqpi}).
The phase factors ${\cal Z}_{\mu\nu}$ in (\ref{GamCR})
are given by
\beq
{\cal Z}_{\mu\nu} = \ee^{2\pi i q \, \varepsilon _{\mu\nu}/L} \ .
\label{Zmunu}
\eeq

\subsection{matrix-field correspondence}
\label{mat-field}

In this subsection we construct a map from fields to matrices
associated with the introduction of space-time noncommutativity.
Generally on the lattice,
the noncommutativity parameter $\Theta_{\mu\nu}$ and the 
period matrix $\Sigma_{\mu\nu} = \ell \delta_{\mu\nu} $ 
should satisfy certain restrictions,
which go away in the continuum limit \cite{AMNS2,AMNS3}.
In the particular example discussed here, 
this amounts to requiring that $b = \pm 2$,
which we assume in what follows\footnote{Eq.(4.11) of \cite{AMNS3},
which is needed for the consistency of the noncommutative algebra of
discretized coordinates, requires that $b$ be even.
The restriction explained before eq.(4.22) of \cite{AMNS3},
which is necessary for a proper discretization of the star-product,
furthermore requires that $b = \pm 2$.
Eq.(4.38) of \cite{AMNS3}, which is related to a property
of star-gauge invariant observables, is satisfied automatically
in the present case.}.
This requires in particular that $L$ is odd.

The plane wave $\ee ^{2 \pi i m_\mu x_\mu / \ell}$,
where $m_\mu$ is a $D$-dimensional integer vector,
is mapped to $N \times N$ unitary matrices
\beq
J_{\vec{m}} \defeq
\left[\prod_{ \mu=1}^D\left(Z _ \mu\right)^{m_ \mu}\right]
{}~\ee ^{- \pi i\sum_{ \mu< \nu}\Theta_{ \mu\nu}m_ \mu m_ \nu} \ ,
\label{defJ}
\eeq
where the phase factor $\ee ^{- \pi i\sum_{ \mu< \nu}
\Theta_{ \mu\nu}m_ \mu m_ \nu}$
is included so that
\beq
J_{-\vec{m}}=(J_{\vec{m}}) ^ \dag \ .
\label{Jconjg}
\eeq
Since $(Z_{\mu})^{L}=1$
the matrix $J_{\vec{m}}$ is periodic with respect to
$m_\mu$ with period $L$.
Arbitrary complex-valued functions $\phi(x)$ 
on the periodic lattice $\Lambda _\ell$
can be mapped to a matrix by first Fourier transforming
and then replacing $\ee ^{2 \pi i m_\mu x_\mu / \ell}$ by $J_{\vec{m}}$.
It follows that $\phi (x)$ is mapped to 
\beq
\Phi  = \frac{1}{N^2}\sum _{x \in  \Lambda_\ell } \phi(x) 
\Delta (x) \ ,
\label{complete}
\eeq
where the $N \times N$ matrices $\Delta (x)$ are defined as
\beq
\Delta (x) \defeq  \frac{1}{N^2} 
\sum _{\vec{m}} 
J_{\vec{m}}~\ee ^{ 2 \pi i   m_\mu x_\mu / \ell } \ .
\label{defDelta}
\eeq
Here the summation for $\vec{m}$ is taken over
a Brillouin zone $-\frac{L-1}{2} \le m_\mu \le \frac{L-1}{2}$.

Note that $\Delta (x)$ is hermitian 
due to the property (\ref{Jconjg}).
It is periodic with respect to $x_\mu$ with period $\ell$.
It is easy to check that $\Delta (x)$ 
possesses the following properties.
\beqa
\label{trDelta}
\Tr \Delta (x) &=& N  \\
\label{sumDelta}
\sum _{x_\mu \in \Lambda_\ell } \Delta (x) &=& N^2 {\bf 1}_N \\
\frac{1}{N} \Tr \Bigl[ \Delta  (x)  \Delta (y) \Bigr] &=& 
N^2 \delta_{x,y} \ .
\label{orthogonal}
\eeqa
Due to eq.~(\ref{orthogonal}), one can invert (\ref{complete}) as
\beq
\phi(x)= \frac{1}{N}\,\Tr\Bigl[\Phi \,\Delta(x)\Bigr] \ .
\label{invert}
\eeq
Therefore, there is actually a one-to-one correspondence between
matrices and fields in the present case.
The number of degrees of freedom matches exactly
due to (\ref{tildep0def});
the matrix has $N^2$ elements and the corresponding field 
depends on $L^D$ space-time points.
Note also that using (\ref{trDelta}) with (\ref{complete}) or
using (\ref{sumDelta}) with (\ref{invert}),
one obtains
\beq
\frac{1}{N}\Tr \Phi  = 
\frac{1}{N^2}\sum _{x \in \Lambda_\ell } \phi(x)  \ .
\label{trace}
\eeq
Summing a function $\phi(x)$ over the torus corresponds 
to taking the trace of the corresponding matrix $\Phi$.


The product of two fields $\phi _1(x)$ and $\phi _2(x)$ 
on the noncommutative torus
should be defined through the product of corresponding matrices
$\Phi_1$ and $\Phi_2$ obtained by the map (\ref{complete}).
This
defines the so-called ``star-product''
\beq
\phi_1(x) \star \phi_2(x)
\defeq
\frac{1}{N} \Tr [\Phi_1 \Phi_2 \Delta (x) ] \ .
\label{stardef}
\eeq
Using the definition of $\Delta (x)$,
the star-product (\ref{stardef}) can be 
written explicitly in terms of $\phi_\alpha (x)$ as
\beq
\phi_1(x) \star \phi_2(x)
= \frac{1}{N^2} \sum _{y \in \Lambda_\ell }
\sum _{z \in \Lambda_\ell }
 \phi_1(y) \phi_2 (z)
\ee ^{-2 i (\theta ^{-1})_{\mu\nu}  
( x _\mu - y_\mu ) ( x _\nu - z_\nu )} \ ,
\label{starexplicit}
\eeq
where $\theta _{\mu\nu}$ is the dimensionful noncommutativity parameter 
given by
\beq
\theta _{\mu\nu} = 
\frac{1}{2\pi} \ell ^2 \Theta_{\mu\nu}
= - \frac{1}{2\pi} b L \latsp ^2 \varepsilon_{\mu\nu}
\label{dimfultheta}
\eeq

In the continuum, the star-product is usually written as
\beq
\phi_1(x) \star \phi_2(x) = \phi_1(x) \exp 
\Bigl( i \frac{1}{2} \theta_{\mu\nu}
\overleftarrow{\del_\mu}
\overrightarrow{\del_\nu}
\Bigr) \phi_2(x)  \ ,
\label{starprod_cont}
\eeq 
which can be rewritten in an integral form as
\beq
\phi_1(x) \star \phi_2(x) = 
\frac1{\pi ^D|\det\theta |}\,\int\!\!\!\int\dd ^D y ~\dd ^D z ~
\phi _1 (y)\, \phi _2 (z)~
\ee ^{- 2 i (\theta^{-1})_{\mu\nu} (x-y)_\mu (x-z)_\nu }  \ .
\label{starprod_cont_kernel}
\eeq
One can easily check that the star-product (\ref{starexplicit}) 
is a proper discretized version of (\ref{starprod_cont_kernel}).
The discretized star-product 
(\ref{starexplicit}) enjoys all the algebraic properties
of the continuum star-product (\ref{starprod_cont}).
In particular, 
\beq
\sum _{x_i \in \Lambda_\ell } 
\phi_1(x) \star \phi_2(x) \star \cdots \star \phi_n(x)
\eeq
is invariant under cyclic permutations of $\phi_\alpha (x)$
and 
\beq
\sum _{x_i \in \Lambda_\ell } 
\phi_1(x) \star \phi_2(x) 
= \sum _{x_i \in \Lambda_\ell } 
\phi_1(x) \phi_2(x)  \ .
\label{bilin}
\eeq


Finally let us mention the properties of the shift operator
$\Gamma_\mu$, which 
is expressed in terms of matrices
as (\ref{twisteat}). From (\ref{Zprimelatticecomm2}) it follows that
\beq
\Gamma_\mu \Delta (x ) \Gamma_\mu  ^\dag 
= \Delta ( x - \latsp \hat{\mu}) \ ,
\eeq
which implies
\beq
\phi (x + \latsp \hat{\mu}) =
\frac1N\,\Tr\Bigl[\Gamma_\mu \Phi \Gamma_\mu  ^\dag 
\Delta(x )\Bigr] \ .
\eeq
Let us consider the field $S_\mu (x)$, which corresponds
to the matrix $\Gamma_\mu$.
\beq
S_\mu  (x) = \frac{1}{N}\,\Tr\Bigl[\Gamma _\mu 
\,\Delta(x)\Bigr] 
= \exp \left(i 2 \pi \frac{L-1}{2} \varepsilon _{\mu\nu} 
\frac{x _\nu}{\ell} \right)  \ .
\label{defS}
\eeq
Such a field has the property
\beq
S_\mu (x) \star \phi (x) \star S_\mu (x) ^* = 
\phi (x + \latsp \hat{\mu}) \ ,
\label{shiftS}
\eeq
for an arbitrary field $\phi (x)$.
Obviously this property is quite peculiar to 
noncommutative geometry.
It plays an important role in the construction of chiral gauge theories
as we shall see in Section \ref{chiral}.

\subsection{noncommutative Yang-Mills theories on the lattice}
\label{NCG}

Let us define noncommutative U($r$) Yang-Mills theory
on the lattice.
As in commutative lattice gauge theories,
we introduce the link variables $U_\mu (x)$,
which can be regarded as 
$r \times r $ matrix fields on the periodic lattice $\Lambda _\ell$.
The unitarity condition of the link variables should be naturally replaced
by
\beq
U_\mu (x) ^\dag \star U_\mu (x)
= U_\mu (x) \star U_\mu (x)^\dag = \id _r 
\mbox{~~~~~(no summation over $\mu$)} \ ,
\label{starunitary}
\eeq
which means that the matrix field $U_\mu (x)$ is `star-unitary'.
Note that the matrix $U_\mu (x)$ for a given $x$ is not necessarily unitary.
Let us introduce $n \times n$ matrices $\hat{U}_\mu$, where $n= rN$, by
\beq
\hat{U} _\mu  = \frac{1}{N^2}\sum _{x \in \Lambda_\ell } 
 U_\mu (x) \otimes \Delta (x)  \ .
\label{mavsfu}
\eeq
Due to (\ref{starunitary}), the matrix $\hat{U}_\mu$ 
should be unitary.
One can invert (\ref{mavsfu}) by using (\ref{orthogonal}) as
\beq
U_\mu (x)= \frac{1}{N}\,\Tr  \Bigl[\hat{U} \,
 \Delta(x)
\Bigr] \ ,
\label{invertU}
\eeq
where we note that 
the trace $\Tr$ is taken over $N$-dimensional indices corresponding
to the `space-time coordinates' only.
Therefore, there is a one-to-one correspondence between
$r \times r $ star-unitary matrix field $U_\mu (x)$ 
and $n \times n$ unitary matrix $\hat{U}$.

One can write down the action for the noncommutative Yang-Mills theory
as
\beq
S _{\rm NCYM} = -\beta \epsilon ^ D \,
\sum _{x\in \Lambda_\ell} \sum_{\mu\neq\nu} \tr  \Bigl[ 
U_\mu (x) \star
U_\nu (x + \latsp \hat{\mu}) \star
U_\mu (x + \latsp \hat{\nu}) ^\dag \star
U_\nu (x) ^\dag \Bigr] \ ,
\label{latticeaction}
\eeq
where the trace $\tr $ is taken over the gauge indices.
The action (\ref{latticeaction})
is invariant under the star-gauge transformation
\beq
U_\mu (x)\mapsto g(x) \star U_\mu (x)
\star g(x+\latsp \hat{\mu})^\dag  \ ,
\label{latticestargaugetr}
\eeq
where the gauge function $g(x)$ 
is star-unitary, $g(x)\star g(x)^{\dag}=
g(x)^{\dag} \star g(x) = \id_r$.

Let us rewrite the action (\ref{latticeaction}) 
in terms of the matrices.
Defining the unitary matrices $\hat{V}_\mu$ as
\beq
\hat{V}_\mu = \hat{U}_\mu 
\Gamma_\mu 
\ ,
\eeq
where $\Gamma_\mu$ are the SU($N$) matrices
defined by (\ref{twisteat}), we arrive at 
\beq
S_{\rm NCYM} = - N \beta
\,\sum_{\mu\neq\nu} {\cal Z}_{\mu\nu}~\Tr \tr
\Bigl(\hat{V}_\mu\,\hat{V}_\nu\,
\hat{V}_\mu^\dag\,\hat{V}_\nu^\dag\Bigr) \ ,
\label{EKaction}
\eeq
where the phase factor ${\cal Z}_{\mu\nu}$ is given 
by (\ref{Zmunu}) \footnote{Throughout the paper,
we use $\Tr$ for 
the trace over $N$-dimensional indices corresponding
to the `space-time coordinates',
and $\tr$ for the trace over $r$-dimensional gauge indices.}.
The action (\ref{EKaction})
is nothing but the twisted Eguchi-Kawai model \cite{EK,GO}.
The star-gauge invariance (\ref{latticestargaugetr})
of (\ref{latticeaction}) corresponds to the SU($n$) invariance
\beq
\hat{V}_\mu \, \mapsto \, \hat{g}\,  \hat{V}_\mu \, \hat{g}^\dag \ ,
\label{SUsym}
\eeq
of the action (\ref{EKaction}), 
where $\hat{g}$ is an SU($n$) matrix.

The integration measure for the U($n$) matrices
$\hat{U}_\mu$ of the twisted Eguchi-Kawai model 
is given by the Haar measure of the U($n$) group,
which respects the SU($n$) invariance (\ref{SUsym}).
This naturally defines
a star-gauge invariant measure for the star-unitary matrix field 
$U_\mu (x)$.
Thus the twisted Eguchi-Kawai model can be interpreted as
a noncommutative U($r$) Yang-Mills theory on the periodic lattice.

Finally, we comment on the continuum limit $\latsp \rightarrow 0$.
We recall that 
the dimensionful noncommutativity parameter $\theta_{\mu\nu}$
is given by (\ref{dimfultheta}).
Therefore, in order to obtain a finite value of $\theta _{\mu\nu}$,
we have to take the large $N(=L^{D/2})$ limit simultaneously
with the continuum limit $\latsp \rightarrow 0$ fixing $L \latsp ^2$.
Note that such a large $N$ limit is different from the one
needed to reproduce ordinary large $N$ gauge theories
from twisted Eguchi-Kawai models \cite{GO},
where one takes the large $N$ limit first for fixed $\latsp$
followed by the continuum limit $\latsp \rightarrow 0$.
(In either large $N$ limit, the coupling constant $\beta$ in
(\ref{EKaction}) should be tuned properly as a function of $\latsp$.)
Existence of the new large $N$ limit with fixed $L \latsp ^2$
has been observed by Monte Carlo simulation 
in $D=2$ \cite{NakajimaNishimura}.
In this limit, the size of the torus $\ell = \latsp L$
goes to infinity.
In order to have both $\theta _{\mu\nu}$ and $\ell$ finite,
one has to use a more general construction \cite{AMNS1,AMNS2,AMNS3}.

\setcounter{equation}{0}
\section{Chiral fermions on the noncommutative torus}
\label{chiral}

In this section, we incorporate chiral fermions 
in the lattice noncommutative gauge theory 
described in the previous section.
If one introduces fermions naively, one encounters the 
doubling problem as in ordinary lattice gauge 
theories \footnote{The fermion doubling problem is addressed in 
Refs.~\cite{Balachandran:2000qu,Gracia-Bondia:1998za,Lizzi:1997vr}
in the context of noncommutative geometry,
and in Refs.~\cite{Kitsunezaki:1998iu,Sochichiu:2000fs}
in matrix models.}.
This makes the implementation of chiral fermions nontrivial.


In Ref.~\cite{Kitsunezaki:1998iu}, the Eguchi-Kawai model
was used to define a unitary matrix version
of the IIB matrix model,
which describes a toroidal compactification of the target space
of type IIB superstrings.
Chiral fermion without species doublers has been 
implemented with the formalism of Ref.~\cite{NN},
which is actually equivalent to the more recent 
approach \cite{Luescher_abelian},
except for the choice of the fermion measure.
In particular, the implementation maintains
the SU($n$) invariance (\ref{SUsym}).
Recalling that the star-gauge invariance of noncommutative gauge theories
corresponds to the SU($n$) invariance (\ref{SUsym}) of the
twisted Eguchi-Kawai model,
it is suggested that we can implement 
chiral fermions in noncommutative gauge theories with manifest
star-gauge invariance.
We will see this explicitly in what follows.

\subsection{lattice Dirac fermions with manifest chiral symmetry}
\label{latDirac}

Let us introduce Dirac fields $\psi (x)$, $\bar{\psi} (x)$
on the periodic lattice $\Lambda _\ell$.
In the case of the fundamental representation,
the fields $\psi (x)$ and $\bar{\psi} (x)$ are 
$r$-component column and row vectors, respectively,
in the internal space,
and transform under star-gauge transformation 
(\ref{latticestargaugetr}) as
\beq
\psi (x) \mapsto g(x) \star \psi (x) ~~~~~;~~~~~
\bar{\psi} (x) \mapsto \bar{\psi} (x) \star g(x)^\dag \ .
\label{fundaferm}
\eeq
We introduce the forward and backward covariant derivative 
on the noncommutative torus as
\beqa
\nabla _\mu \psi &=& 
\frac{1}{\latsp}
\Bigl[ U_\mu (x) \star \psi (x+\latsp \hat{\mu}) - \psi (x) \Bigr]\n
\nabla _\mu ^* \psi &=& 
\frac{1}{\latsp}
\Bigl[\psi (x) - U_\mu (x-\latsp \hat{\mu} )^\dag 
\star \psi (x-\latsp \hat{\mu}) \Bigr]  \ .
\eeqa
One can define an action for Dirac fermion as
\beq
S_{\rm w} = \epsilon ^D
\sum _{x} \bar{\psi} (x) \, \star D_{\rm w} \, \psi (x) \ ,
\label{WilsonDirac}
\eeq
where $D_{\rm w}$ is the Wilson-Dirac operator given as
\beq
D_{\rm w}  = \frac{1}{2}
\sum _{\mu = 1} ^D
\{ \gamma _\mu (\nabla _\mu ^ * + \nabla _\mu  ) + 
\latsp  \nabla _\mu ^ *  \nabla _\mu \} \ .
\label{WDop}
\eeq
The star ($\star$) written explicitly in (\ref{WilsonDirac})
--but not the ones hidden in the operator $D_{\rm w}$--
can be omitted due to the property (\ref{bilin}).
For trivial gauge configuration $U_\mu (x)= \id_r$,
the action (\ref{WilsonDirac}) agrees with the 
usual Wilson-Dirac action in the commutative space-time.
The O($\latsp$) term in (\ref{WDop}) is the Wilson term,
which is introduced to give the unwanted species doublers 
masses of O($\latsp^{-1}$).
However, as is well known, the Wilson term breaks chiral symmetry.

In the case of the adjoint representation,
the Dirac fields are $r \times r$ matrices in the internal space,
and transform under star-gauge transformation 
(\ref{latticestargaugetr}) as
\beq
\psi (x) \mapsto g(x) \star \psi (x) \star g(x)^\dag  ~~~~~;~~~~~
\bar{\psi} (x) \mapsto g(x) \star \bar{\psi} (x) \star g(x)^\dag \ .
\label{adjferm}
\eeq
The forward and backward covariant derivative 
can be defined as
\beqa
\nabla _\mu \psi &=& 
\frac{1}{\latsp}
\Bigl[ U_\mu (x) \star \psi (x+\latsp \hat{\mu}) \star U_\mu (x)^\dag
- \psi (x) \Bigr]\n
\nabla _\mu ^* \psi &=& 
\frac{1}{\latsp}
\Bigl[\psi (x) - U_\mu (x-\latsp \hat{\mu} )^\dag 
\star \psi (x-\latsp \hat{\mu})
\star U_\mu (x-\latsp \hat{\mu} ) \Bigr]  \ .
\eeqa
One can define an action for Dirac fermion as
\beq
S_{\rm w} = \epsilon ^D
\sum _{x} \tr \Bigl[ \bar{\psi} (x) \, \star D_{\rm w} \, \psi (x) 
\Bigr] \ .
\label{WilsonDirac2}
\eeq

As in the commutative case \cite{ML},
we can maintain an exact chiral symmetry
on the lattice without doublers by 
using a Dirac operator which satisfies the Ginsparg-Wilson relation.
\beq
\gamma _ 5 D +  D \gamma _5  = \latsp \, D \gamma _5 D \ .
\label{GWrelation}
\eeq
Assuming $D^{\dag}=  \gamma_{5} D \gamma_{5}$,
which is usually refered to as ``$\gamma_{5}$-Hermiticity'',
we can define the unitary operator $\hat{\gamma}_5$
\beq
\hat{\gamma}_5 = \gamma_5 (1 - \latsp D) \ ,
\label{gamhatdef}
\eeq
which has the properties
\beqa
\label{gamhatsquare}
(\hat{\gamma}_5)^2 &=& 1 \\
\gamma_5 D &=& - D \hat{\gamma}_5 \label{gamhatproj} \ .
\eeqa
Eq.\ (\ref{gamhatproj}) implies that
the corresponding action
\beqa
S &=& \epsilon ^D \sum_{x} \bar{\psi}(x) \star D \psi (x) 
\mbox{~~~~~~~~~for fundamental representation}\\
S &=& \epsilon ^D \sum_{x} \tr \Bigl[ \bar{\psi}(x) \star D \psi (x) \Bigr]
\mbox{~~~~~~~~~for adjoint representation} 
\label{GWaction}
\eeqa
has the symmetry
\beq
\psi (x) \mapsto \ee ^{i \alpha \hat{\gamma} _5 }  \psi (x) ~~~~~;~~~~~
\bar{\psi} (x) \mapsto \bar{\psi} (x) \ee ^{i \alpha \gamma _5 } \ ,
\eeq
which is the exact lattice chiral symmetry.

As in the commutative case \cite{overlapDirac},
an explicit solution to the Ginsparg-Wilson relation (\ref{GWrelation}),
which transforms covariantly under star-gauge transformation, 
can be given by
\beq
D = \frac{1}{\latsp} \Bigl\{ 1 - {A ( A ^ \dag A )^{-{1\over 2}}} \Bigr\} \ ,
~~~~~A =  1 - \latsp D_{\rm w}  \ ,
\label{overlapDirac}
\eeq
where the Wilson-Dirac operator $D_{\rm w}$ is defined 
by eq.\ (\ref{WDop}).

\subsection{projecting out chiral fermions}
\label{chiralproj}

Due to the property (\ref{gamhatproj}),
we can project a chiral fermion out of Dirac fermion
as in \cite{Luescher_abelian} by imposing the constraint
\beqa
\label{gammahat_projection}
\hat{\gamma}_5 \, \psi (x) &=& \psi (x) \\
\bar{\psi} (x) \, \gamma_5 &=& - \bar{\psi} (x) \ .
\label{gammahat_projection2}
\eeqa
In order to define the integration measure for $\psi (x)$
and $\bar{\psi}(x)$ after the projection (\ref{gammahat_projection})
and (\ref{gammahat_projection2}),
we take a complete orthogonal basis 
$\{ \varphi _j (x) ; j=1, \cdots , K\}$ 
and $\{ \bar{\varphi} _j (x) ; j=1, \cdots , \bar{K}\}$
for the solutions to (\ref{gammahat_projection}) and
(\ref{gammahat_projection2}) respectively as
\beqa
\label{projphi}
\hat{\gamma}_5 \, \varphi _j (x) = \varphi _j (x)
 ~~~~~&;&~~~~~(\varphi _j  ,  \varphi _k ) \defeq
\sum _{x \in \Lambda _\ell } \varphi _j (x) ^\dag \, \varphi _k (x) 
= \delta_{jk}     \\
 \bar{\varphi} _j (x)  \, \gamma_5 = - \bar{\varphi} _j (x)
 ~~~~~&;&~~~~~(\bar{\varphi} _j  ,  \bar{\varphi} _k ) = \delta_{jk} \ .
\label{projphibar}
\eeqa
The general solution to (\ref{gammahat_projection})
and (\ref{gammahat_projection2})
can be written in terms of the complete basis as
\beq
\psi (x) = \sum_j c_j  \, \varphi_j (x) ~~~~~;~~~~~
\bar{\psi} (x) = \sum_j \bar{c}_j  \, \bar{\varphi}_j (x)
\eeq
where the coefficients $c_j$ and $\bar{c}_j$ are Grassmann variables.
Then the integration measure for the chiral fermion fields
can be defined by 
\beq
{\cal D} \psi  \, {\cal D} \bar{\psi}  = \prod_j \dd c_j  \cdot
\prod_k \dd \bar{c}_k \ .
\label{measure}
\eeq
Here we recall that unlike the usual chirality operator,
$\hat{\gamma}_5$ depends on the gauge configuration $U_\mu (x)$.
Therefore, the basis $\{ \varphi _j (x) \} $ has to be specified
for each gauge configuration $U_\mu (x)$,
whereas one can take the same basis $\{ \bar{\varphi} _j (x) \} $ 
for all gauge configurations.
If we choose a different basis $\{ \varphi _j (x) \}$,
the integration measure defined by (\ref{measure}) 
may change by a phase factor which depends on the gauge configuration.
Therefore, the crucial question that arises here is whether one can fix
this gauge-field dependent phase ambiguity
in such a way that the integration measure is star-gauge invariant.

We fix the gauge-field dependent phase ambiguity
in the following way.
First we specify the complete basis 
for the gauge configuration $U_\mu  ^{(0)} (x)$
as $\{\varphi_j^{(0)} (x)\}$.
Then for a general gauge configuration $U_\mu (x)$, 
we require that the basis $\{ \varphi _j (x) \}$
satisfy the condition
\beq
\det_{j,k} \Bigl\{ ( \varphi_j^{(0)} , \varphi_k ) \Bigr\} 
\mbox{~:~real positive} \ .
\label{WBcondition}
\eeq
Although this does not determine the basis $\{ \varphi _j (x)\} $ uniquely,
it determines the integration measure uniquely
(up to a constant phase factor)
as far as the determinant is non-zero.
This phase choice was originally proposed in Ref.~\cite{NN}
for commutative lattice gauge theory, where 
the `reference configuration' $U_\mu  ^{(0)}(x)$ 
was taken to be $U_\mu  ^{(0)}(x)=\id _r$.
However, the corresponding fermion measure is {\em not} gauge invariant,
due to the fact that the configuration $U_\mu  ^{(0)}(x)=\id_r$
is not invariant under the gauge 
transformation\footnote{In Ref.~\cite{NN}, it was speculated that
the gauge averaging will lead to a sensible definition of 
a chiral gauge theory when the fermion content satisfies
the anomaly cancellation condition. The 
two-dimensional U(1) case, where analytic results
are available, has been studied intensively in \cite{IzubuchiNishimura},
where a particular class of gauge anomaly,
which may contribute significantly in the gauge averaging,
has been identified. See also Ref.~\cite{Golterman} for a clarification 
on this point.}.

In the noncommutative space-time, there exists a gauge configuration
$\tilde{U}_\mu (x)$ 
which is invariant under arbitrary star-gauge transformations.
\beq
g(x) \star \tilde{U}_\mu  (x) \star g(x+\latsp \hat{\mu}) ^\dag 
= \tilde{U}_\mu  (x) \ .
\label{tildeU}
\eeq
The unique solution to (\ref{tildeU}) --up to a constant phase--
is $\tilde{U}_\mu  (x) = S_\mu (x)^{*} \id _r$, where $S_\mu (x)$
is the field defined in (\ref{defS}),
and the invariance (\ref{tildeU}) follows 
from the property (\ref{shiftS}).
We are going to use this configuration
$\tilde{U}_\mu  (x)$ as the reference configuration;
namely $U_\mu  ^{(0)}(x) = \tilde{U}_\mu  (x)$.

Let us recall that the dimension $K$ of the solution
space of (\ref{gammahat_projection}) depends on the gauge configuration.
Here we focus on configurations $U$ for which $K[U] = \bar{K}$,
namely we restrict ourselves to the topologically trivial sector
of gauge configurations (i.e., there is no baryon number violation).
In order for the condition (\ref{WBcondition}) to work,
we should have $K[U] = K[U ^{(0)}]$, which requires
$K[U  ^{(0)}]= \bar{K}$.
In fact, this is guaranteed for the adjoint representation,
but not for the fundamental representation.
Let us consider the adjoint representation first.
For the particular configuration $U_\mu (x) = U_\mu  ^{(0)}(x)$,
one finds that the covariant derivatives
$\nabla _\mu$ and $\nabla _\mu ^* $ vanish 
and hence we obtain
\footnote{From this, it also follows that we can use 
the usual chiral basis as 
$\{\varphi_j^{(0)} (x)\}$ in (\ref{WBcondition}).
In fact, the basis $\{\varphi_j^{(0)} (x)\}$
is all we need to specify the fermion measure.
The particular configuration $\tilde{U}_\mu  (x)$, which is
not a smooth function of $x$ (See eq.\ (\ref{defS})),
is used for convenience in the proof of the star-gauge invariance,
but it does not play any role in specifying the measure.}
$\hat{\gamma}_5 =\gamma_5$,
which means in particular that $K[U ^{(0)}] = \bar{K}$ as announced.
Since the determinant in (\ref{WBcondition}) is nonzero 
for generic configurations in the topologically trivial sector,
the condition (\ref{WBcondition}) works in the adjoint representation.
In the case of the fundamental representation, however, 
the covariant derivatives $\nabla _\mu$ and $\nabla _\mu ^* $ 
do not vanish for the configuration $U_\mu (x) = U_\mu  ^{(0)} (x)$
and there is no reason for $K[U ^{(0)}] = \bar{K}$.
In fact we have checked numerically that $K[U ^{(0)}] \neq \bar{K}$ 
in general.
Therefore, the condition (\ref{WBcondition}) does not work 
in the fundamental representation.

\subsection{star-gauge invariance of the fermion measure}
\label{proof}

For the reason mentioned in the previous subsection,
we restrict ourselves to the adjoint representation in what follows.
Let us prove that the corresponding integration measure is indeed 
star-gauge invariant.
Here we denote the basis as $\varphi_j [U]$,
to express the gauge field dependence manifestly
and suppress the space-time argument $x$.
Then
\beqa
g \star \varphi _j [U]  \star g^\dag
&=& \sum _k \varphi_k [U^g] \, Q_{kj}[U,g] \n
g \star \varphi_j^{(0)} \star g^\dag
&=& \sum _k \varphi_k ^{(0)} \, Q_{kj}^{(0)}[g]  \ ,
\eeqa
where $U^g$ represents the star-gauge transformed configuration
and $Q_{kj}$ and $Q_{kj}^{(0)}$ are {\it unitary} transformation matrices.
It follows that
\beqa
\det_{j,k} \{ ( \varphi_j^{(0)} , \varphi_k(U) ) \}  
& =& \det _{j,k}\Bigl\{ 
( g \star \varphi_j^{(0)} \star g^\dag , 
g \star \varphi _k [U] \star g^\dag  ) \Bigr\} \n
&=&  \det _{j,k}
\Bigl\{ \sum _{m,n}
(Q^{(0)}_{mj})^* ( \varphi_m ^{(0)}  , \varphi_n [U^g] ) Q_{nk}
\Bigr\} \n
&=&  (\det Q^{(0)})^*  \cdot \det _{j,k} 
\Bigl\{ ( \varphi_j^{(0)}  , \varphi _k [U^g] ) \Bigr\} \cdot  \det Q \ .
\eeqa
Due to the condition (\ref{WBcondition}), 
we obtain $\det Q = \det Q^{(0)}$.
Writing the transformed fermion field $\psi ' = g \star \psi \star g^\dag $
in two different ways
\beqa
\psi ' &=& \sum _{j} c_j ' \, \varphi _j [U^g]  \n 
\psi ' &=& \sum_j  c_j  \, g \star \varphi_j [U] \star g^\dag 
= \sum_{j,k}  c_j  \,  \varphi_k [U^g] \,  Q_{kj} \ ,
\eeqa
we obtain the transformation of the coefficients $c_j$ as
\beq
 c_j  ' = \sum _k Q_{jk} \, c_k  \ .
\eeq
Therefore, the measure for $\psi$ fields transforms as
\beq
{\cal D} \psi ' 
= \prod_j \dd c_j ' = (\det Q[U,g])^{-1} \prod_j \dd c_j 
= (\det Q^{(0)} [g])^{-1} \, {\cal D} \psi   \ .
\eeq
Similarly, the measure for $\bar{\psi}$ fields transforms as
\beq
{\cal D} \bar{\psi} ' = 
 (\det \bar{Q} [g])^{-1} \, {\cal D} \bar{\psi}  \ ,
\eeq
where the unitary matrix $\bar{Q} _{kj} [g]$ is defined by
\beq
 g \star \bar{\varphi} _j \star g ^\dag = \sum _k \bar{\varphi}_k \, 
\bar{Q} _{kj} [g] \ .
\eeq
It is easy to check that $\det Q^{(0)}[g] = (\det \bar{Q}[g])^*$ 
for arbitrary $g$, from which it follows that
the fermion measure (\ref{measure}) is star-gauge invariant.

\subsection{chiral determinants}
\label{chiraldeterminants}

Let us consider the fermionic partition function defined as
\beq
Z[U] = \int {\cal D} \psi  \, {\cal D} \bar{\psi} \, 
\ee ^{- S} \ ,
\eeq 
where the action and the measure is given by
(\ref{GWaction}) and (\ref{measure}) respectively.
One can write it explicitly as
\beq
Z[U] = 
\det _{j,k} \Bigl\{ ( \bar{\varphi}_j  , \varphi _k [U] ) \Bigr\}  \ ,
\label{defZ}
\eeq
with the condition (\ref{WBcondition}).
This quantity actually vanishes identically
due to the fact that the constant modes
$\psi (x) \propto \id _r$ and
$\bar{\psi} (x) \propto \id _r$
do not appear in the action. (Recall that we are considering
the adjoint representation.)

In order to define a nonzero partition function
we have to insert these modes as external lines as
\beq
Z'[U] = \int {\cal D} \psi  \, {\cal D} \bar{\psi} \, 
\prod _{\alpha=p/2+1}^{p} \Bigl\{
 \sum_x 
\tr \Bigl[ \bar{\psi}_\alpha (x) \Bigr] \Bigr\}
\,  \prod _{\beta=1}^{p/2} \Bigl\{
\sum_y \tr \Bigl[ \psi_\beta (y) \Bigr] \Bigr\} \, 
\ee ^{- S} \ ,
\eeq 
where the integer $p$ is the dimension of the (Dirac) spinor space
$p = 2 ^{D/2}$ and 
we took the Weyl representation $\gamma_5 = \mbox{diag} (1,\cdots , 1,
-1 ,  \cdots , -1)$.
The partition function $Z'[U]$ can be written in the same
form as (\ref{defZ}) with the same condition (\ref{WBcondition}),
but the bases in (\ref{defZ}) and (\ref{WBcondition})
should now be constructed in the subspace orthogonal
to the direction of the constant modes $\psi (x) \propto \id _r$ and
$\bar{\psi} (x) \propto \id _r$.
The partition function $Z'[U]$ is nonzero for generic configurations,
and moreover it is invariant under the star-gauge transformation:
\beq
Z'[U^g] = Z'[U] \ .
\eeq

The fermion determinant $Z'[U]$ can be calculated 
numerically for each gauge configuration $U$ in the following way.
In order to obtain the basis $\{ \varphi _j [U] \} $,
it is convenient to note that
\beq
\hat{\gamma}_5 = {\gamma_5 A (A ^ \dag A)^{-{1\over 2}}} 
= \frac{H}{\sqrt{ H ^ \dag H } } \ ,
\eeq
where $H$ is a hermitian operator defined as
\beq
H = \gamma_5 A = \gamma_5 (1- \epsilon D_{\rm w}) \ .
\eeq
Then instead of using (\ref{projphi}), one can construct
the basis $\{ \varphi _j [U] \} $ as the eigenvectors
of the hermitian operator $H$
\beq
H \varphi _j = E_j \varphi _j
\label{overlapHam}
\eeq
with positive eigenvalues $E_j > 0$ \cite{NN}.
Actual calculations become very simple by rewriting
equations such as (\ref{overlapHam}) using the matrix-field correspondence
described in Section \ref{mat-field}.
We calculate the fermion determinant $Z'[U]$ in the $D=2$, U(1) case.
The parameters in (\ref{apbqpi}) are chosen to be
$b=2$, $s=-1$ and $q = \frac{L+1}{2}$.
For $L=3,5,7,9,11$, we have checked that $Z'[U]$ has a nontrivial
phase, which is invariant under random star-gauge transformations.

\setcounter{equation}{0}
\section{Gauge anomalies in the continuum calculations}
\label{continuum}

In this Section, 
we will review some relevant aspects of gauge anomalies in 
noncommutative chiral gauge theories and compare our results
on the lattice to some continuum calculations. Gauge anomalies in 
noncommutative chiral gauge theories has been extensively studied 
in the literature recently, mostly restricted to four dimensions 
\cite{Gracia-Bondia:2000pz,Bonora:2000he,Martin:2000qf,Ardalan:2000qk}.

One of the intriguing features of anomalies in noncommutative chiral gauge theories is the
fact that the anomaly cancellation condition differs from the one encountered in the commutative 
version of the theory, yet it is independent of the noncommutativity parameter. 
This can be seen as a consequence of UV/IR mixing \cite{iruv} reflecting the 
noncommutativity between 
the two limits $\theta_{\mu\nu}\rightarrow 0$ and $\Lambda_{\rm UV}\rightarrow \infty$.
In computing the anomalous Ward identities for the gauge current in noncommutative chiral 
gauge theories one recovers 
the ordinary commutative anomaly by taking the commutative limit 
while keeping the ultraviolet cutoff fixed; once the the ultraviolet cutoff is removed the 
limit $\theta_{\mu\nu}\rightarrow 0$ does not retrieve the commutative Ward identity anymore.

\subsection{generalities}

We consider noncommutative U($r$) gauge theory
in the continuum.
The gauge field $A_\mu (x)$ is a $r \times r$ hermitian
matrix field, which transform as
\beq
\delta_{\eta} A_\mu  = \del_\mu \eta  
- i  (A_\mu \star \eta - \eta \star A_\mu) \ ,
\eeq
where the $r \times r$ hermitian matrix field $\eta (x)$ represents 
the gauge function.
For the fermion fields $\psi$, $\bar{\psi}$ transforming in the (anti) 
fundamental representation of U($r$) as
\begin{eqnarray}
\delta_{\eta}\psi = i\eta\star\psi
&~~~;~~~&
\delta_{\eta}\bar{\psi} = - i\bar{\psi} \star \eta \nonumber 
\\
\delta_{\eta}\psi = -i\psi\star \eta 
&~~~;~~~&
\delta_{\eta}\bar{\psi}  =  i \eta  \star \bar{\psi}   \ ,
\label{fund}
\end{eqnarray}
the covariant derivatives are given respectively by
\begin{eqnarray}
D _\mu \psi
&=& \partial_{\mu}\psi-i A_{\mu}\star\psi\,\,,
\nonumber \\
D _\mu \psi&=& \partial_{\mu}\psi+i \psi
\star A_{\mu}\,\ .
\label{fc}
\end{eqnarray}
In the U(1) case, (anti) fundamental fermions have respectively charge $\pm 1$ 
with respect to the gauge field;
any other charge assignment would spoil the covariance of (\ref{fc}) 
under (\ref{fund}). 
For the fermion fields transforming in the adjoint representation as
\begin{equation}
\delta_{\eta}\psi = i(\eta\star\psi-\psi\star\eta) 
~~~;~~~
\delta_{\eta}\bar{\psi} = i(\eta\star\bar{\psi}-\bar{\psi}\star\eta) \ ,
\label{adj}
\end{equation}
the covariant derivative is given  by
\begin{equation}
D_\mu \psi=\partial_{\mu}\psi - i (A_{\mu}\star\psi-\psi\star A_{\mu}) \ .
\label{acoup}
\end{equation}

\subsection{review of the four-dimensional case}

When $D=4$ the calculation of the gauge anomaly in noncommutative chiral gauge theories with fermions 
in the (anti) fundamental representation has been done using either the heat-kernel method 
\cite{Gracia-Bondia:2000pz} or the perturbative analysis \cite{Martin:2000qf}, leading to 
the following anomaly cancellation condition
\begin{equation}
\tr \,\left(T^{a}T^{b}T^{c}\right)=0,
\label{ccond}
\end{equation}
where $T^{a}$ are the generators of the representation of the gauge group.
Notice that this condition is much stronger that the one for commutative chiral gauge theories, 
$\tr \, \left(T^{a}\{T^{b},T^{c}\}\right)=0$. 
One particular feature of noncommutative chiral gauge theories with the (anti) fundamental coupling 
is that, unlike the case of their commutative counterparts, the form of eq. (\ref{ccond}) 
implies that the only way to cancel the anomaly by adding several 
types of fermions is to have 
the same number of fermions of one chirality 
transforming in the fundamental and
antifundamental representations 
(or equivalently the same number of left- and right-handed fermions
with the same transformation properties). This is easy to see when 
the gauge group is U(1) \cite{Hayakawa:2000yt} and 
the anomaly cancellation condition reduces
to the one for commutative chiral QED, $\sum_{i}q_{i}^3=0$. Now, however, since the 
charges are fixed by the transformation properties of the fermion the only way to cancel the 
total anomaly is by considering pairs of fermions transforming respectively in the fundamental 
and the antifundamental representation (or again, pairs of fermions with opposite chirality
and the same U(1) charge). The resulting theory is therefore vector-like.

The situation changes when the fermions transform in the adjoint representation
of U($r$). A direct calculation of the triangle anomaly 
shows that noncommutative chiral gauge theories 
with adjoint fermions are anomaly free 
\cite{Martin:2000qf,Ardalan:2000qk}. This is not so surprising, though, since
four-dimensional chiral fermions in the adjoint representation can be alternatively 
formulated as Majorana fermions. Our result in the previous Section
using the lattice construction of the noncommutative chiral gauge theory is 
consistent with these observations.

\subsection{the two-dimensional case}

The analysis of Section \ref{chiral} is valid in any even dimension and therefore
noncommutative chiral gauge theories with adjoint fermions can be constructed
on the lattice without breaking of star-gauge invariance for any $D=2k$.
Thus it would be interesting to see whether the cancellation of anomalies for adjoint fermions 
in the continuum found in four dimensions also happens
in other dimensions different from four. The simplest case is a gauge theory with 
adjoint chiral fermions in Euclidean two-dimensional noncommutative space-time. In two 
dimensions the analog of the triangle diagram is the chiral fermion loop with two gauge field 
insertions (see Fig. 1). Because we are in two dimensions the chirality matrix 
$\gamma_{5}$ satisfies the identity $\gamma_{\mu}\gamma_{5}=i\epsilon_{\mu\nu}\gamma_{\nu}$. 
Therefore we can write\footnote{We use the conventions $\gamma_{5}=i\gamma_{0}\gamma_{1}$, 
$\gamma_{\mu}^{\dagger}=\gamma_{\mu}$.}
\begin{equation}
\gamma_{\mu}{\rm P}_{\pm}={1\over 2}(\delta_{\mu\nu}\pm i\epsilon_{\mu\nu})\gamma_{\nu}\,,
\end{equation}
where ${\rm P}_{\pm}={1\over 2}(1\pm\gamma_{5})$ are the projectors over right(left)-handed chiralities. 
As a consequence the amplitude  with chiral couplings to the gauge field in Fig. 1, 
$\Gamma_{\mu\nu}^{ab}(p^2)_{+}$, is related to the contribution of the same 
diagram with vector-like couplings $\Gamma_{\mu\nu}^{ab}(p^2)$ by
\begin{equation}
\Gamma_{\mu\nu}^{ab}(p^2)_{+}={1\over 2}\left(\delta_{\mu\alpha}+i\epsilon_{\mu\alpha}
\right)\Gamma^{ab}_{\alpha\nu}(p^2)\,.
\end{equation}
\begin{figure}
\centerline{\epsfxsize=2.5truein \epsffile{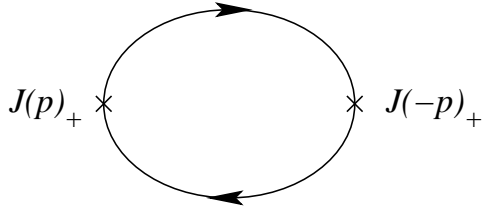}}
\caption{Feynman diagram contributing to the two-dimensional gauge anomaly.}
\end{figure}

The term $\Gamma_{\mu\nu}^{ab}(p^2)$ has been computed in \cite{Moreno:2000xu,Moreno:2001kt} 
where it was seen that it only receives contributions from the planar part of the amplitude.
Using that result we find that the chiral anomaly is given by
\begin{equation}
p_{\mu}\Gamma_{\mu\nu}^{ab}(p^2)_{+}={i\over 4\pi}(d^{ace}d^{bce}+
f^{ace}f^{bce})\epsilon_{\mu\nu}p_{\mu}\,.
\label{anom2d}
\end{equation}
Here $f^{abc}=-2i \, \tr \, (T^{a}[T^{b},T^{c}])$ are the structure constants 
and $d^{abc}=2 \, \tr  \, (T^{a}\{T^{b},
T^{c}\})$ are the anomaly coefficients of 
the gauge group (with the normalization $\tr  \, (T^{a}T^{b}) =
{1\over 2}\delta^{ab}$). 

For U($r$) a simple calculation leads to  
$d^{ace}d^{bce}+f^{ace}f^{bce}=2r\delta^{ab}$ and we find the following 
anomalous Ward identity
\begin{equation}
p_{\mu}\langle J_{\mu}^{a}(p)_{+}\,J_{\nu}^{b}(-p)_{+}\rangle = {ir\over 2\pi}p_{\mu}
\epsilon_{\mu\nu}\delta^{ab},
\label{nv}
\end{equation}
where $J_{\mu}^{a}(p)_{+}$ is the chiral gauge current derived from the Lagrangian 
${\cal L}= \tr \, \left(\bar{\psi}\star \gamma _\mu D_\mu
\,{\rm P}_{+}\psi\right)$ with the covariant derivative 
defined in (\ref{acoup}). Notice that, as in the four-dimensional case, even if the anomalous 
Ward identity (\ref{anom2d}) is independent of the noncommutativity parameter the actual form of the anomaly 
differs from the one corresponding to a commutative chiral gauge theory, where the
term $d^{ace}d^{bce}$ would be absent from the group theoretical factor. 
Again this is a consequence of UV/IR mixing. However, when the gauge group is U($r$) with $r>1$, the
actual numerical value of the prefactor on the right-hand side of (\ref{nv}) coincides with the 
one for the corresponding commutative gauge theory. This is due to the U($r$) identity $f^{ace}f^{bce}=
d^{ace}d^{bce}=r\delta^{ab}$. Thus, the contribution from the $d$-symbols compensates the factor ${1\over 2}$ 
that comes from the fact that only the planar part contributes to the diagram in Fig. 1 
\footnote{Alternatively, this can be seen by noticing that for U($r$) ($r>1$) all dependence
on $\theta$ in the calculation of the anomaly disappears from the very beginning since 
$d^{ace}d^{bce}\sin^2{{1\over 2}\theta(p,q)}+f^{ace}f^{bce}\cos^2{{1\over 2}\theta(p,q)}=r\delta^{ab}$.}.

In the case when the gauge group is U(1), in order to connect with the literature \cite{renorma},
it is convenient to change the normalization of the only generator from $T^{0}={1\over \sqrt{2}}$ 
to $T^{0}=1$. Thus, the anomaly can be obtained from eq. (\ref{anom2d}) by setting $f^{000}=0$ and
$d^{000}=2$. The interesting thing about this case is that the naive $\theta_{\mu\nu} \rightarrow
0$ limit in the classical action
gives a free theory, whereas the calculation of the anomaly in the noncommutative case
gives a $\theta$-independent nonvanishing result which resembles that of a two-dimensional 
nonabelian commutative gauge theory (cf. \cite{renorma,Moreno:2001kt}).

We have seen that in the case of two-dimensional noncommutative chiral gauge theories with fermions 
in the adjoint representation there is a gauge anomaly. One way to cancel this anomaly 
is by adding a second adjoint fermion with opposite chirality, with 
Dirac operator 
$\gamma_\mu D_\mu {\rm P}_{-}$.
As a consequence, the resulting theory 
will be nonchiral and will satisfy the vector Ward identity $p_{\mu}\langle J_{\mu}^{a}(p)
J_{\nu}^{b}(-p)\rangle \equiv p_{\mu}\Gamma_{\mu\nu}^{ab}(p^2)=0$, with $J_{\mu}^{a}(p)=
J_{\mu}^{a}(p)_{+}+J_{\mu}^{a}(p)_{-}$. 
This perturbative result in the continuum contrasts with our construction on the lattice where
gauge invariance is preserved without renouncing to the chiral character of the theory. 

Finally, the calculation for the case of fermions in the (anti) fundamental representation
goes along similar lines as the one outlined above. In this case the net noncommutative phase
in the diagram is equal to one, and the anomaly can be obtained 
by replacing the group theoretical factor in (\ref{anom2d}) with $2\, \tr \, 
(T^{a}T^{b})=\delta^{ab}$.
Again for the U(1) case it is convenient to change the normalization to $T^{0}=1$. Then the
resulting anomaly is half the value found for the adjoint U(1) case.

\setcounter{equation}{0}
\section{Summary and discussion}
\label{summary}

In this paper, we have studied noncommutative chiral gauge theories
on the lattice by considering a chiral fermionic action containing a Dirac 
operator satisfying the Ginsparg-Wilson relation. 
When the fermions are in the adjoint representation of the gauge group U($r$) we 
found that one can construct chiral gauge theories on the lattice
with manifest star-gauge invariance in arbitrary even dimensions.
In the continuum, it was known that the gauge anomaly cancels 
for the adjoint representation in four dimensions.
However, if we do the same continuum calculation in two dimensions,
the gauge anomaly remains. Therefore, our results indicate a certain 
dependence on the regularization procedure.

This is quite reminiscent of the situation in commutative space-time with odd dimensions.
There the parity anomaly conflicts with the gauge symmetry in some cases \cite{oddD}.
If one imposes the gauge invariance, one obtains the parity anomaly.
If one imposes parity invariance, one obtains the gauge anomaly.
The results depend on regularization schemes.
But the results are related to each other by adding an appropriate
local counterterm, which is nothing but a Chern-Simons term.
It is therefore suggested that, in the 
case at hand, the gauge anomaly obtained in the continuum
calculation might be cancelled by adding some counterterm as well.
Although the noncommutativity of the space-time inevitably introduces
non-locality, we speculate that the counterterm is written
in such a way that the non-locality is restricted to that encoded in the
star-product.
It would be interesting to identify the explicit form of the
counterterm and we leave it for future investigations.


\section*{Acknowledgments}
We would like to thank L.~Alvarez-Gaum\'e, J.L.F.~Barb\'on, S.~Iso and H.~Kawai 
for helpful discussions and D.H. Adams for correspondence on his work \cite{Adams:2000yi}. 
M.A.V.-M. is supported by 
EU Network ``Discrete Random Geometry'' Grant HPRN-CT-1999-00161, ESF 
Network no. 82 on ``Geometry and Disorder'', Spanish Science Ministry Grant AEN99-0315 
and University of the Basque Country Grants UPV 063.310-EB187/98 and UPV 172.310-G02/99.

\end{document}